\documentclass[a4paper]{cas-sc}

\usepackage[authoryear]{natbib}
\usepackage{mathrsfs}

\newcommand{\tr}{\operatorname{tr}}
\newcommand{\dev}{\operatorname{dev}}
\newcommand{\C}{\mathbf{C}}
\newcommand{\F}{\mathbf{F}}
\newcommand{\Sigmaln}[1][]{\boldsymbol{\Sigma}_{\mathrm{ln}#1}}
\newcommand{\Sigmalndot}{\boldsymbol{\dot{\Sigma}}_{\ln}}
\newcommand{\ubalpha}{\boldsymbol{\alpha}}
\newcommand{\ubbeta}{\boldsymbol{\beta}}
\newcommand{\ubsigma}{\boldsymbol{\sigma}}
\newcommand{\ubtau}{\boldsymbol{\tau}}
\newcommand{\ulambda}{\lambda}
\newcommand{\umu}{\mu}
\newcommand{\ueta}{\eta}



\graphicspath{{figures/}}

\begin{document}
\let\WriteBookmarks\relax
\def\floatpagepagefraction{1}
\def\textpagefraction{.001}

\shorttitle{A finite-strain logarithmic viscoelastic model for Antarctic ice shelves based on an additive split}
\shortauthors{M. Nutte et~al.}

\title[mode=title]{A finite-strain logarithmic viscoelastic model for Antarctic ice shelves based on an additive split}

\author[1]{Maxime Nutte}[orcid=0009-0008-5541-6907]
\cormark[1]
\ead{maxime.nutte@uct.ac.za}
\credit{Methodology, Software, Investigation, Writing -- original draft}

\author[1]{Sebastian Skatulla}
\credit{Conceptualization, Supervision, Writing -- review \& editing}

\author[2]{Carlo Sansour}
\credit{Conceptualization, Methodology}

\affiliation[1]{organization={Polar Engineering Research Group, University of Cape Town},
            city={Cape Town},
            country={South Africa}}

\affiliation[2]{organization={Bethlehem University},
            city={Bethlehem},
            country={Palestine}}

\cortext[1]{Corresponding author}

\begin{abstract}
Ice shelves lose mass primarily by calving, a process controlled by the near-front stress field on timescales that span elastic flexure and viscous creep. We formulate a finite-strain Maxwell model for glacier ice in logarithmic strain space. The Hencky strain of a fixed reference configuration is split additively at the level of rates into elastic and viscous parts; the spring is isotropic Hencky elasticity and the dashpot is a Glen-type power law written on the logarithmic strain rate and its work-conjugate stress. At infinitesimal strain the dashpot coincides with Glen's flow law; the elastic strains in the ice-shelf configurations of this paper remain in that regime. The model is integrated with a midpoint evaluation of \eqref{eq:Cdot} and a backward-Euler correction of the trial dual, and implemented in a finite-element setting. After a viscoelastic column benchmark, the formulation is applied to an idealised ice tongue, including depth-dependent density and moduli, temperature-dependent fluidity, and cliff geometries with a frontal foot or basal undercutting. The resulting stress fields show how viscoelasticity and front morphology control tension near the terminus.
\end{abstract}


\begin{keywords}
finite strain \sep logarithmic strain \sep viscoelasticity \sep Glen's flow law \sep ice shelves \sep calving
\end{keywords}

\maketitle

\section{Introduction}

Ice shelves are the floating extensions of the grounded ice sheet. They remain attached at a grounding line, spread under their own weight, and terminate at a calving front whose position controls the horizontal extent of much of the Antarctic Ice Sheet. Calving remains among the least well constrained components of large-scale ice-sheet models \citep{scambos2009,christmann2019}, which typically either hold the ice front fixed or parametrize its motion through a phenomenological rate. A physics-based description of that rate has to start from the stress and strain that develop near the terminus.

\begin{figure}[htb!]
\centering
\includegraphics[width=0.62\linewidth]{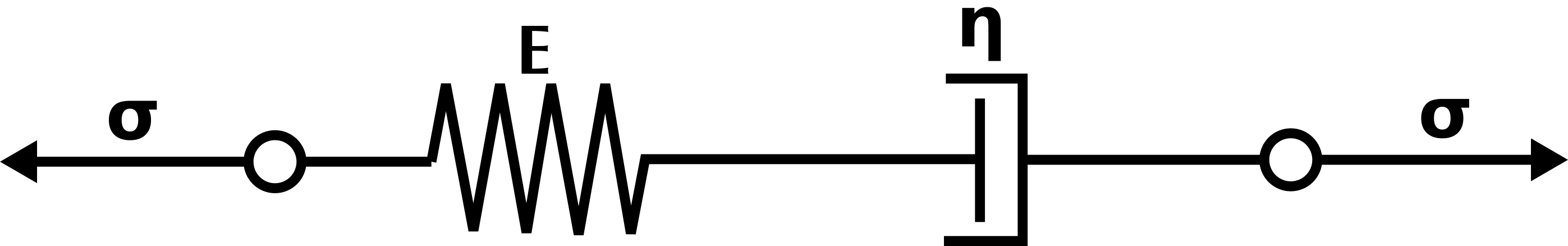}
\caption{Maxwell element: elastic spring in series with a viscous dashpot. The two branches carry the same stress; the total strain is the sum of the elastic and viscous parts.}
\label{fig:maxwell}
\end{figure}

Ice-sheet models treat ice as a viscous fluid governed by Glen’s flow law and solve for velocities \citep{glen1955,greve2009,macayeal1989,cuffey2010}. The description is appropriate for creep over years to millennia, but it omits the elastic response. Near a calving front, ocean swell, tidal flexure, and the sudden traction-free condition after a calving event act on timescales of seconds to days, on which ice also deforms elastically \citep{christmann2016,christmann2019}. In those regimes a purely viscous Glen fluid is expected to underestimate the tensile stress that can open crevasses. Both the instantaneous elastic deformation and the long-term Glen creep must therefore be retained if that tensile field is to be computed across a calving event and the subsequent hold. The rheological model typically used to capture both is a Maxwell model (Fig.~\ref{fig:maxwell}): the total strain rate is the sum of an instantaneous elastic contribution and a viscous contribution that recovers Glen creep \citep{christmann2016,christmann2019}. The Maxwell time (the ratio of viscosity to shear modulus) is of the order of hours for temperate ice at typical shelf stresses \citep{christmann2016}, so elasticity and creep occupy distinct but overlapping windows of the glaciological record.

At the terminus itself, this tensile field is generated by the local buoyancy equilibrium, even in a still ocean \citep{reeh1968}. Ice overburden increases linearly with depth, while hydrostatic seawater pressure acts only on the submerged part of the front and does not fully balance the weight of the ice. Because these forces have different distributions and lines of action, their mismatch generates a net bending moment that tends to deform an unfooted vertical cliff. Finite-element calculations of viscous and viscoelastic fronts have long shown that this bending produces a bell-shaped tensile maximum near the upper surface, typically about half an ice thickness inland \citep{fastook1982,christmann2016}. Cliff geometry modifies the same bending moment. A submerged ice foot can reverse the direction of deflection and reduce the surface tensile maximum (the ``footloose'' mechanism associated with rampart--moat profiles \citep{wagner2014,mosbeux2020}) whereas an undercut face shortens the hydrostatic lever arm, concentrates tension, and shifts the peak inland of the crest \citep{christmann2016}. Neither geometric effect is a perturbation of the far-field flow. The same bending response is further modified by variations in ice properties through the thickness. Firn densification and a vertical temperature gradient make density, stiffness and fluidity depth-dependent, thereby changing both the buoyancy mismatch and the local Maxwell time. A vertical viscosity gradient has been shown to produce an internal bending moment even without a submerged bench \citep{buck2024,glazer2026}, although its net effect on the surface tensile peak depends on the particular profiles. These near-front effects control small-scale calving at relatively homogeneous fronts, as distinct from rift-driven tabular break-off \citep{benn2018,christmann2019}.

Over the months to years between those calving events, linearized strain is sufficient only while the accumulated stretch remains small \citep{christmann2019}. Horizontal spreading of an ice tongue increases strain even when strain rates stay modest, so that the small-strain hypothesis fails although the rates themselves may never become large. Comparing a Maxwell model written on the infinitesimal strain with its finite-strain counterpart, \citet{christmann2019} found that stress and strain maxima already differ by about $5\%$ after one year and by about $30\%$ after a decade. A finite-strain formulation reaches a critical stress or strain sooner, so that a calving criterion based on either quantity would predict a higher rate; a Glen-type viscous branch raises the stress further relative to a constant viscosity. Linear viscoelasticity therefore remains useful on the short timescales of a single swell event or a tidal cycle, but finite kinematics become necessary once the hold between calving events, or a sequence of such events, is measured in years to decades \citep{schroeder2026}. Ice tongues and ice cliffs do undergo finite geometry change. The elastic strains themselves stay small (of order $10^{-5}$ at typical shelf stresses, and at most of order $10^{-4}$ at near-front tensile peaks), so the finite-strain setting is required by the accumulated viscous stretch and by the evolving geometry of the front, not by a large elastic stretch. The calculations below address that near-front stress field, without a prescribed calving rate or a propagating crack.

Meeting that finite-strain requirement, \citet{schroeder2026} constructed a Maxwell--Glen model for ice from the multiplicative split \(\F=\F_e\F_v\). The viscous factor is the internal variable of an intermediate configuration and is integrated by an exponential map that preserves isochoric flow. That scheme is a return mapping of multiplicative type, and assembling the consistent tangent is then laborious, as the algorithmic moduli must be transported through \(\F_e\). The formulation has been verified on a self-weighted column and on a hydrostatic ice-shelf benchmark.

Those features motivate a logarithmic realisation of the same Maxwell--Glen rheology. The Hencky strain \(\ubalpha\) of a fixed reference configuration is split additively at the level of rates \(\dot{\ubalpha}=\dot{\ubalpha}_e+\dot{\ubalpha}_{in}\), without \(\F_v\) \citep{sansour2001,sansour2001cmame,sansour2003}; the dual stress is work-conjugate on that configuration. The formulation is motivated by the following features.
\begin{itemize}
\item Ice-sheet models are rate-based: they solve Glen's law for velocity \citep{gagliardini2013,larour2012}. A Glen-type dashpot written on the logarithmic rate remains a constitutive statement of the same type, and therefore couples to those solvers without a change of kinematic variables.
\item The additive structure of the formulation makes its extension to anisotropic behaviour straightforward, unlike a multiplicative decomposition \(\F=\F_e\F_v\), for which that extension can become involved, owing to objective rates and the transport of tensors through an intermediate configuration.
\item Time integration retains the structure of the small-strain Maxwell scheme, while the kinematics remain finite.
\end{itemize}
To our knowledge, such an additive logarithmic Maxwell formulation has not previously been applied to glacier ice. After the range of the additive logarithmic split has been examined, a column benchmark against \citet{schroeder2026} precedes an application to an idealised ice tongue with stratified density and moduli, Arrhenius fluidity, and alternative cliff geometries.

Section~\ref{sec:kin} sets out the kinematics and work conjugacy of the logarithmic strain, together with the evaluation of the exponential map. Section~\ref{sec:const} states the Maxwell--Glen-type model, the time integrator, and the algorithmic tangent. Section~\ref{sec:validity} examines the range of the additive logarithmic formulation. Section~\ref{sec:examples} presents a representative numerical example comparing the additive reduction at $m=1$ with the column problem under axial compression discussed in \citet{schroeder2026}. 

\section{Kinematics and work conjugacy}
\label{sec:kin}

\subsection{Logarithmic strain and work conjugacy}
\label{sec:logstrain}

Let $\mathscr{B}\subset\mathbb{R}^3$ define a body. A motion of $\mathscr{B}$ is represented by a one-parameter mapping $\boldsymbol{\varphi}_t:\mathscr{B}\to\mathscr{B}_t$, where $t\in\mathbb{R}$ is the time and $\mathscr{B}_t$ is the current configuration at time $t$. Associated with each material point of the body are the position vectors $\mathbf{X}\in\mathscr{B}$ at the reference configuration and $\mathbf{x}\in\mathscr{B}_t$ at the current configuration. One has $\boldsymbol{\varphi}_t(\mathbf{X})=\mathbf{x}$. Explicit reference to $t$ will be omitted. The displacement is $\mathbf{u}=\mathbf{x}-\mathbf{X}$. The tangent map related to $\boldsymbol{\varphi}$ is the deformation gradient $\F$,
\begin{equation}
\label{eq:FJC}
\F
\;=\;
\mathrm{Grad}\,\boldsymbol{\varphi}
\;=\;
\mathbf{I}+\mathrm{Grad}\,\mathbf{u},
\qquad
J=\det\F>0.
\end{equation}
With the help of $\F$ one defines the right Cauchy--Green tensor
\begin{equation}
\label{eq:C}
\C=\F^{\mathrm{T}}\F.
\end{equation}
In addition, $\F$ possesses the polar decomposition $\F=\mathbf{R}\mathbf{U}$, with $\mathbf{U}$ symmetric positive definite and $\mathbf{R}$ a rotation. By the symmetry of $\mathbf{U}$ and $\C$, and the fact that $\det\mathbf{U}>0$, the logarithmic strain $\ubalpha$ can be introduced, following \citet{sansour2001}, according to
\begin{equation}
\label{eq:alpha}
\ubalpha
\;=\;
\ln\mathbf{U}
\;=\;
\tfrac12\ln\C.
\end{equation}
Equally, $\mathbf{U}=\exp\ubalpha$, and with $\ubbeta=2\ubalpha$,
\begin{equation}
\label{eq:Cexp}
\C
\;=\;
\exp\ubbeta
\;=\;
\mathbf{I}+\ubbeta+\frac{\ubbeta^2}{2!}+\frac{\ubbeta^3}{3!}+\cdots.
\end{equation}
Because $\ubalpha$ is an isotropic function of $\C$, it shares the eigenframe of $\mathbf{U}$. When $\C\simeq\mathbf{I}$, $\ubalpha$ reduces to the infinitesimal strain tensor. The trace identity
\begin{equation}
\label{eq:tralpha}
\tr\ubalpha=\ln J,
\end{equation}
or equivalently $\exp(\tr\ubalpha)=J$, follows at once from \eqref{eq:alpha}.

Let a superposed dot denote the material time derivative. From \eqref{eq:Cexp} one has $\C=\exp\ubbeta$, which leads to
\begin{equation}
\label{eq:Cdotexp}
\dot{\C}
\;=\;
\frac{\partial\exp\ubbeta}{\partial\ubbeta}:\dot{\ubbeta},
\end{equation}
thus
\begin{equation}
\label{eq:CdotA}
\dot{\C}
\;=\;
\mathbb{A}:\dot{\ubbeta},
\end{equation}
with
\begin{equation}
\label{eq:Adef}
\mathbb{A}
\;=\;
\frac{\partial\exp\ubbeta}{\partial\ubbeta}
\end{equation}
a fourth-order tensor, evaluated in Section~\ref{sec:expmap}. Hence
\begin{equation}
\label{eq:betadot}
\dot{\ubbeta}
\;=\;
\mathbb{A}^{-1}:\dot{\C},
\end{equation}
and, with \(\ubbeta=2\ubalpha\), the logarithmic rate
\begin{equation}
\label{eq:A}
\dot{\ubalpha}
\;=\;
\tfrac12\,\mathbb{A}^{-1}:\dot{\C}.
\end{equation}
At small strain, \(\dot{\ubalpha}\) coincides with the infinitesimal strain rate.
We thus define \(\dot{\C}\) by differentiating \eqref{eq:C},
\begin{equation}
\label{eq:Cdot}
\dot{\C}
\;=\;
\dot{\F}^{\mathrm{T}}\F+\F^{\mathrm{T}}\dot{\F}.
\end{equation}

Let $\ubsigma$ be the Cauchy stress tensor, $\ubtau=J\ubsigma$ the Kirchhoff stress, and $\mathbf{S}$ the second Piola--Kirchhoff stress, with $\ubtau=\F\mathbf{S}\F^{\mathrm{T}}$ and $\mathbf{P}=\F\mathbf{S}$ the first Piola--Kirchhoff stress. The specific internal power is written on the material pair \((\mathbf{S},\dot{\C})\),
\begin{equation}
\label{eq:power}
\mathscr{W}
\;=\;
\tfrac12\,\mathbf{S}:\dot{\C},
\end{equation}
using \eqref{eq:Cdot} and without passing through the spatial stretching. The dual of $\ubalpha$ is obtained from \eqref{eq:Cexp} without any assumption of isotropy. Expanding $\C=\exp\ubbeta$ in \eqref{eq:power} gives
\begin{equation}
\label{eq:Wexp}
\begin{aligned}
\mathscr{W}
&\;=\;
\tfrac12\,\mathbf{S}:\dot{\C}\\
&\;=\;
\Bigl(
\mathbf{S}
+\tfrac{1}{2!}\bigl(\mathbf{S}\ubbeta+\ubbeta\mathbf{S}\bigr)
+\tfrac{1}{3!}\bigl(\mathbf{S}\ubbeta^2+\ubbeta\mathbf{S}\ubbeta+\ubbeta^2\mathbf{S}\bigr)
+\cdots
\Bigr):\dot{\ubalpha}.
\end{aligned}
\end{equation}
Hence the dual variable of the logarithmic strain reads
\begin{equation}
\label{eq:Sigln}
\Sigmaln
\;=\;
\mathbf{S}
+\tfrac{1}{2!}\bigl(\mathbf{S}\ubbeta+\ubbeta\mathbf{S}\bigr)
+\tfrac{1}{3!}\bigl(\mathbf{S}\ubbeta^2+\ubbeta\mathbf{S}\ubbeta+\ubbeta^2\mathbf{S}\bigr)
+\cdots,
\end{equation}
so that $\mathscr{W}=\Sigmaln:\dot{\ubalpha}$. Due to the symmetry of $\mathbf{S}$ and $\ubbeta$, $\Sigmaln$ itself is symmetric. Inverting the pairing \eqref{eq:A} yields the work-conjugate reconstruction
\begin{equation}
\label{eq:S}
\mathbf{S}
\;=\;
\mathbb{A}^{-1}:\Sigmaln.
\end{equation}
Cauchy stress follows by $\ubsigma=J^{-1}\F\mathbf{S}\F^{\mathrm{T}}$. If $\Sigmaln$ commutes with $\C$, \eqref{eq:S} reduces to $\mathbf{S}=\C^{-1}\Sigmaln$.

\subsection{Evaluation of the exponential map and its derivative}
\label{sec:expmap}

Following \citet{sansour1998}, $\mathbb{A}$ is computed from \eqref{eq:Cexp} by the Cayley--Hamilton reduction of powers of $\ubbeta$ of order three and higher to the span of $\mathbf{I}$, $\ubbeta$ and $\ubbeta^2$. Let the principal invariants of $\ubbeta$ be
\begin{equation}
\label{eq:invbeta}
I_1=\tr\ubbeta,
\qquad
I_2=\tfrac12\bigl(I_1^2-\tr\ubbeta^2\bigr),
\qquad
I_3=\det\ubbeta,
\end{equation}
with
\begin{equation}
\label{eq:dinvbeta}
\frac{\partial I_1}{\partial\ubbeta}=\mathbf{I},
\qquad
\frac{\partial I_2}{\partial\ubbeta}=I_1\mathbf{I}-\ubbeta,
\qquad
\frac{\partial I_3}{\partial\ubbeta}=I_3\ubbeta^{-1}.
\end{equation}
Powers of order $n\ge 3$ then admit the representation
\begin{equation}
\label{eq:betan}
\ubbeta^n
=\gamma_0^{(n)}\,\mathbf{I}
+\gamma_1^{(n)}\,\ubbeta
+\gamma_2^{(n)}\,\ubbeta^2,
\end{equation}
where the scalar coefficients obey, for $n\ge 4$,
\begin{equation}
\label{eq:gammarec}
\gamma_0^{(n)}=I_3\,\gamma_2^{(n-1)},
\qquad
\gamma_1^{(n)}=\gamma_0^{(n-1)}-I_2\,\gamma_2^{(n-1)},
\qquad
\gamma_2^{(n)}=\gamma_1^{(n-1)}+I_1\,\gamma_2^{(n-1)},
\end{equation}
starting from $\gamma_0^{(3)}=I_3$, $\gamma_1^{(3)}=-I_2$, $\gamma_2^{(3)}=I_1$. Hence
\begin{equation}
\label{eq:expphi}
\exp\ubbeta
=\varphi_0\,\mathbf{I}
+\varphi_1\,\ubbeta
+\varphi_2\,\ubbeta^2,
\end{equation}
with
\begin{equation}
\label{eq:phi}
\begin{aligned}
\varphi_0
&=1+\frac{1}{3!}\,I_3
+\sum_{n=4}^{\infty}\frac{1}{n!}\,\gamma_0^{(n)},
\\
\varphi_1
&=1-\frac{1}{3!}\,I_2
+\sum_{n=4}^{\infty}\frac{1}{n!}\,\gamma_1^{(n)},
\\
\varphi_2
&=\frac12+\frac{1}{3!}\,I_1
+\sum_{n=4}^{\infty}\frac{1}{n!}\,\gamma_2^{(n)}.
\end{aligned}
\end{equation}
Differentiating \eqref{eq:expphi} with respect to $\ubbeta$ yields
\begin{equation}
\label{eq:Aexp}
\begin{aligned}
\mathbb{A}
&=
\frac{1}{3!}\frac{\partial I_3}{\partial\ubbeta}\otimes\mathbf{I}
+\sum_{n=4}^{\infty}\frac{1}{n!}\frac{\partial\gamma_0^{(n)}}{\partial\ubbeta}\otimes\mathbf{I}
-\frac{1}{3!}\frac{\partial I_2}{\partial\ubbeta}\otimes\ubbeta
+\sum_{n=4}^{\infty}\frac{1}{n!}\frac{\partial\gamma_1^{(n)}}{\partial\ubbeta}\otimes\ubbeta\\
&\quad
+\frac{1}{3!}\frac{\partial I_1}{\partial\ubbeta}\otimes\ubbeta^2
+\sum_{n=4}^{\infty}\frac{1}{n!}\frac{\partial\gamma_2^{(n)}}{\partial\ubbeta}\otimes\ubbeta^2
+\varphi_1\,\mathbb{I}
+\varphi_2\frac{\partial(\ubbeta^2)}{\partial\ubbeta}.
\end{aligned}
\end{equation}
The inverse in \eqref{eq:A} is equivalently $\mathbb{A}^{-1}=\partial\ln\C/\partial\C$.

\section{Constitutive model}
\label{sec:const}

\subsection{Additive decomposition of the logarithmic rate}

An additive decomposition of the logarithmic rate is introduced,
\begin{equation}
\label{eq:split}
\dot{\ubalpha}
\;=\;
\dot{\ubalpha}_e
+\dot{\ubalpha}_v.
\end{equation}
Since \(\ubalpha\) is a strain of the fixed reference configuration, \(\dot{\ubalpha}\) is an ordinary material time derivative. Its additive partition is therefore a split of a single material rate, without an additional objective stress or strain rate. This preserves the additive structure familiar from small-strain viscoelasticity while remaining embedded in the finite-strain kinematics through \(\dot{\C}\) and \(\mathbb{A}\) \citep{sansour2001cmame,sansour2003}.

A further advantage follows from the logarithmic character of \(\ubalpha\). The identity \eqref{eq:tralpha} makes the volumetric change a linear invariant of the same strain measure. An incompressible viscous response can therefore be imposed directly on the viscous branch through
\begin{equation}
\label{eq:trv}
\tr\dot{\ubalpha}_v=0,
\qquad
\tr\ubalpha_e=\ln J,
\end{equation}
so that the entire volume change is carried by the elastic part, while the viscous contribution remains isochoric. This avoids introducing a separate volumetric--deviatoric decomposition of the deformation gradient solely to enforce viscous incompressibility.

In the Maxwell construction considered below, the two branches are driven by the same work-conjugate stress, while the total logarithmic rate is partitioned between the elastic and viscous contributions. The additive split in \eqref{eq:split} thus provides a direct finite-strain counterpart of the small-strain Maxwell structure, with the nonlinear kinematics retained in the constitutive operators rather than in the rate decomposition itself.

\subsection{Elastic constitutive law}

The Maxwell element of Fig.~\ref{fig:maxwell} identifies the two rates in \eqref{eq:split}: the spring carries $\dot{\ubalpha}_e$, the dashpot carries $\dot{\ubalpha}_v$, and both branches share the same stress $\Sigmaln$. The spring is isotropic Hencky elasticity, written directly in rate form,
\begin{equation}
\label{eq:hencky}
\Sigmalndot
\;=\;
\ulambda\,(\tr\dot{\ubalpha}_e)\,\mathbf{I}
+2\umu\,\dev\dot{\ubalpha}_e,
\end{equation}
or equivalently $\Sigmalndot=\mathbb{C}^e:\dot{\ubalpha}_e$ with
\begin{equation}
\label{eq:Ce}
\mathbb{C}^e
\;=\;
\ulambda\,\mathbf{I}\otimes\mathbf{I}
+2\umu\,\mathbb{I}^{\mathrm{dev}},
\qquad
\mathbb{I}^{\mathrm{dev}}
\;=\;
\mathbb{I}-\tfrac13\mathbf{I}\otimes\mathbf{I}.
\end{equation}
The Lam\'e moduli $\ulambda$ and $\umu$ are equivalent to Young's modulus $E$ and Poisson's ratio $\nu$ through
\begin{equation}
\label{eq:lame}
\umu
\;=\;
\frac{E}{2(1+\nu)},
\qquad
\ulambda
\;=\;
\frac{E\nu}{(1+\nu)(1-2\nu)}.
\end{equation}
The bulk modulus $K=\ulambda+\tfrac23\umu$ governs the elastic volumetric stiffness.

\subsection{Glen-type dashpot}

Glen's flow law relates the Eulerian stretching $\mathbf{d}=\mathrm{sym}(\dot{\F}\,\F^{-1})$ to the Kirchhoff stress \citep{glen1955}: in the incompressible, isothermal form,
\begin{equation}
\label{eq:glen}
\dev\mathbf{d}
\;=\;
A\,q_{\tau}^{m-1}\,\dev\ubtau,
\qquad
q_{\tau}=\sqrt{\tfrac12\,\dev\ubtau:\dev\ubtau},
\end{equation}
with Glen exponent $m$, typically $m=3$ for glacier ice \citep{greve2009,cuffey2010}. On the logarithmic pair a corresponding dashpot is obtained by assuming that the same power-law structure can be written on the viscous logarithmic rate and its dual. This correspondence is a constitutive assumption rather than a consequence of the finite-strain kinematics. It is exact at infinitesimal strain, where $\dot{\ubalpha}$ coincides with the infinitesimal strain rate and $\Sigmaln$ with $\ubtau$. It remains a close approximation for moderate strains and approximately coaxial loading paths. In that regime the logarithmic dashpot provides a close counterpart to Glen's law while preserving work conjugacy with $\dot{\ubalpha}_v$. With the invariants
\begin{equation}
\label{eq:J2}
J_2
\;=\;
\tfrac12\,\dev\Sigmaln:\dev\Sigmaln,
\qquad
q=\sqrt{J_2},
\end{equation}
and the stress-dependent viscosity
\begin{equation}
\label{eq:eta}
\ueta
\;=\;
\frac{1}{2A\,q^{m-1}}\qquad(q>0),
\end{equation}
the Maxwell identification $\dev\Sigmaln=2\ueta\,\dot{\ubalpha}_v$ leads to the following Glen-\emph{type} flow law
\begin{equation}
\label{eq:glenlog}
\dot{\ubalpha}_v
\;=\;
A\,q^{m-1}\,\dev\Sigmaln.
\end{equation}
The dashpot has no yield threshold and flows for any non-zero $q$. The flow rule is purely deviatoric, and therefore satisfies \eqref{eq:trv}. The dissipation of the dashpot is
\begin{equation}
\label{eq:diss}
\mathscr{D}
=\Sigmaln:\dot{\ubalpha}_v
=\dev\Sigmaln:\dot{\ubalpha}_v
=2A\,q^{m+1}
\ge 0.
\end{equation}
Equality holds only at $q=0$. For $m>1$, the viscosity is formally infinite at $q=0$, while the flow rule itself gives $\dot{\ubalpha}_v=0$.

\subsection{Time integration}
\label{sec:algo}

This section presents the time integration procedure of the constitutive model and details the operations required for the local iterative solution. Let two discrete times $t_n$ and $t_{n+1}$ be given, with increment $\Delta t$. At a material point, $\F_{n+1}$ is known. The kinematic quantities entering the constitutive update are evaluated at the midpoint, while the Maxwell evolution is integrated implicitly over the increment. Accordingly,
\begin{equation}
\label{eq:FCmid}
\F_{n+1/2}
=\tfrac12(\F_n+\F_{n+1}),
\qquad
\dot{\F}_{n+1/2}
=\frac{\F_{n+1}-\F_n}{\Delta t},
\qquad
\C_{n+1/2}
=\F_{n+1/2}^{\mathrm{T}}\F_{n+1/2}
\end{equation}
The rate of $\C$ is that of \eqref{eq:Cdot} at $t_{n+1/2}$,
\begin{equation}
\label{eq:Cdotmid}
\dot{\C}|_{n+1/2}
=\dot{\F}_{n+1/2}^{\mathrm{T}}\F_{n+1/2}
+\F_{n+1/2}^{\mathrm{T}}\dot{\F}_{n+1/2},
\end{equation}
and the logarithmic rate follows from \eqref{eq:A},
\begin{equation}
\label{eq:mid}
\dot{\ubalpha}
=\tfrac12\,\mathbb{A}^{-1}(\C_{n+1/2}):\dot{\C}|_{n+1/2}
\end{equation}
An elastic trial is obtained by freezing the dashpot,
\begin{equation}
\label{eq:aetrial}
\dot{\ubalpha}_e^{\mathrm{trial}}
=\dot{\ubalpha}
\end{equation}
The corresponding trial dual is
\begin{equation}
\label{eq:trial}
\boldsymbol{\dot{\Sigma}}_{\mathrm{ln}}^{\mathrm{trial}}
=\ulambda\,(\tr\dot{\ubalpha}_e^{\mathrm{trial}})\,\mathbf{I}
+2\umu\,\dev\dot{\ubalpha}_e^{\mathrm{trial}},
\qquad
\Sigmaln^{\mathrm{trial}}
=\Sigmaln|_n
+\Delta t\,\boldsymbol{\dot{\Sigma}}_{\mathrm{ln}}^{\mathrm{trial}},
\qquad
\mathbf{s}^{\mathrm{trial}}=\dev\Sigmaln^{\mathrm{trial}}
\end{equation}
The trial second invariant is
\begin{equation}
\label{eq:J2trial}
J_2^{\mathrm{trial}}
=\tfrac12\,\mathbf{s}^{\mathrm{trial}}:\mathbf{s}^{\mathrm{trial}}
\end{equation}
and is therefore fully determined once $\F_n$ and $\F_{n+1}$ are known. With $q_{\mathrm{trial}}=\sqrt{J_2^{\mathrm{trial}}}$, the Glen-type law \eqref{eq:glenlog} is evaluated implicitly at $t_{n+1}$,
\begin{equation}
\label{eq:alphav}
\dot{\ubalpha}_v|_{n+1}
\;=\;
A\,q_{n+1}^{m-1}\,\mathbf{s}_{n+1}
\end{equation}
The dual at $t_{n+1}$ follows from the elastic law on $\dot{\ubalpha}_e=\dot{\ubalpha}-\dot{\ubalpha}_v$,
\begin{equation}
\label{eq:updateexp}
\begin{aligned}
\Sigmaln|_{n+1}
&=\Sigmaln|_n
+\Delta t\bigl[
\ulambda\bigl(\tr(\dot{\ubalpha}|_{n+1}-\dot{\ubalpha}_v|_{n+1})\bigr)\,\mathbf{I}
+2\umu\,\dev(\dot{\ubalpha}|_{n+1}-\dot{\ubalpha}_v|_{n+1})
\bigr]\\
&=\Sigmaln^{\mathrm{trial}}
-2\umu\Delta t\,\dot{\ubalpha}_v|_{n+1},
\end{aligned}
\end{equation}
the last equality because $\tr\dot{\ubalpha}_v=0$. Thus
\begin{equation}
\label{eq:update}
\Sigmaln|_{n+1}
\;=\;
\Sigmaln^{\mathrm{trial}}
-2\umu\Delta t\,\dot{\ubalpha}_v|_{n+1}
\end{equation}
Substituting \eqref{eq:alphav} into \eqref{eq:update} and taking the deviatoric part of both sides gives
$\mathbf{s}_{n+1}=\mathbf{s}^{\mathrm{trial}}-2\umu\Delta t\,A\,q_{n+1}^{m-1}\,\mathbf{s}_{n+1}$.
Hence $\mathbf{s}_{n+1}$ remains collinear with $\mathbf{s}^{\mathrm{trial}}$,
\begin{equation}
\label{eq:sbeta}
\mathbf{s}_{n+1}
=\frac{\mathbf{s}^{\mathrm{trial}}}{1+\beta},
\qquad
\beta
=2\umu\Delta t\,A\,q_{n+1}^{m-1}
\end{equation}
Since $\mathbf{s}_{n+1}$ and $\mathbf{s}^{\mathrm{trial}}$ are collinear, taking their $J_2$-norms yields the scalar residual
\begin{equation}
\label{eq:qres}
q_{n+1}+2\umu\Delta t\,A\,q_{n+1}^{m}
=q_{\mathrm{trial}}
\end{equation}
Thus the local constitutive update reduces to a single scalar nonlinear equation, regardless of the three-dimensional character of the stress and strain tensors. For $q\ge 0$ and $m\ge 1$ the left-hand side is strictly increasing, so a unique root lies in $[0,q_{\mathrm{trial}}]$. It is obtained by Newton iteration,
\begin{equation}
\label{eq:newton}
q^{(k+1)}
=q^{(k)}-\frac{r\bigl(q^{(k)}\bigr)}{r'\bigl(q^{(k)}\bigr)},
\qquad
r(q)=q+2\umu\Delta t\,A\,q^{m}-q_{\mathrm{trial}},
\qquad
r'(q)=1+2\umu\Delta t\,A\,m\,q^{m-1},
\end{equation}
started at $q^{(0)}=q_{\mathrm{trial}}$. For $m>1$, $r$ is convex and increasing on $[0,\infty)$; since $q_{\mathrm{trial}}$ lies to the right of the unique root, Newton's iteration started at $q^{(0)}=q_{\mathrm{trial}}$ converges monotonically from above. For a linear dashpot ($m=1$), $A=1/(2\ueta)$ and \eqref{eq:qres} is solved in closed form,
\begin{equation}
\label{eq:explicit}
\mathbf{s}_{n+1}
=\frac{\mathbf{s}^{\mathrm{trial}}}{1+\Delta t/t_{\mathrm{M}}},
\qquad
t_{\mathrm{M}}=\ueta/\umu,
\end{equation}
which is the standard backward-Euler update for linear Maxwell viscoelasticity and is unconditionally stable. For $m=3$, \eqref{eq:qres} is the depressed cubic
\begin{equation}
\label{eq:qres3}
2\umu\Delta t\,A\,q_{n+1}^{3}+q_{n+1}-q_{\mathrm{trial}}=0,
\end{equation}
or equivalently $q_{n+1}^{3}+p\,q_{n+1}+c=0$ with
\begin{equation}
\label{eq:cardanpc}
p=\frac{1}{2\umu\Delta t\,A}>0,
\qquad
c=-\frac{q_{\mathrm{trial}}}{2\umu\Delta t\,A}.
\end{equation}
Cardano's formula supplies the unique real root
\begin{equation}
\label{eq:cardan}
q_{n+1}
=
\sqrt[3]{-\frac{c}{2}+\sqrt{\frac{c^{2}}{4}+\frac{p^{3}}{27}}}
+
\sqrt[3]{-\frac{c}{2}-\sqrt{\frac{c^{2}}{4}+\frac{p^{3}}{27}}}.
\end{equation}
The discriminant $c^{2}/4+p^{3}/27$ is strictly positive, so there is one real root and a complex-conjugate pair, in agreement with uniqueness for $m>1$, and the expression remains in the reals. Once $q_{n+1}$ is known --- from \eqref{eq:newton} or from \eqref{eq:cardan} --- \eqref{eq:sbeta} and \eqref{eq:update} give the dual.

The stresses at $t_{n+1}$ are reconstructed from \eqref{eq:S},
\begin{equation}
\label{eq:Snp1}
\mathbf{S}_{n+1}
=\mathbb{A}^{-1}(\C_{n+1}):\Sigmaln|_{n+1},
\qquad
\ubsigma_{n+1}
=J_{n+1}^{-1}\F_{n+1}\mathbf{S}_{n+1}\F_{n+1}^{\mathrm{T}}
\end{equation}
If $\Sigmaln|_{n+1}$ commutes with $\C_{n+1}$, the first of \eqref{eq:Snp1} reduces to $\mathbf{S}_{n+1}=\C_{n+1}^{-1}\Sigmaln|_{n+1}$. The hydrostatic part of $\Sigmaln$ remains at its trial value, so the discrete viscous increment is traceless and \eqref{eq:trv} is inherited exactly. Thus, the constitutive state is represented by $\Sigmaln$, while the previous deformation gradient $\F_n$ is retained as part of the time-discrete kinematic history; the viscous strain is not stored.

\subsection{Algorithmic tangent operator}
\label{sec:tangent}

With \eqref{eq:update} at hand, the algorithmic tangent operator can be systematically derived by linearizing $\mathbf{S}$ with respect to $\C$ \citep{sansour2003}. One has first, from \eqref{eq:Snp1} and \eqref{eq:update},
\begin{equation}
\label{eq:Sprod}
\mathbf{S}
=\mathbb{A}^{-1}(\C):\Sigmaln
=\mathbb{A}^{-1}(\C):
\bigl(
\Sigmaln^{\mathrm{trial}}
-2\umu\Delta t\,\dot{\ubalpha}_v
\bigr)
\end{equation}
The derivative with respect to $\C$ is
\begin{equation}
\label{eq:dSdC}
\frac{\partial\mathbf{S}}{\partial\C}
=
\frac{\partial\mathbb{A}^{-1}}{\partial\C}:\Sigmaln
+
\mathbb{A}^{-1}(\C):
\frac{\partial\Sigmaln}{\partial\C}
\end{equation}
The first term is kinematic: $\partial\mathbb{A}^{-1}/\partial\C$, a sixth-order tensor. The second term is constitutive. Let $\Delta\ubalpha=\Delta t\,\dot{\ubalpha}$. Then $\Sigmaln^{\mathrm{trial}}$ in \eqref{eq:trial} depends on $\Delta\ubalpha$ through the elastic law,
\begin{equation}
\label{eq:dtrialC}
\frac{\partial\Sigmaln^{\mathrm{trial}}}{\partial(\Delta\ubalpha)}
=\mathbb{C}^e
=\ulambda\,\mathbf{I}\otimes\mathbf{I}
+2\umu\,\mathbb{I}^{\mathrm{dev}},
\qquad
\frac{\partial\mathbf{s}^{\mathrm{trial}}}{\partial(\Delta\ubalpha)}
=2\umu\,\mathbb{I}^{\mathrm{dev}}
\end{equation}
The viscous correction depends on $q$ through \eqref{eq:sbeta}. Then every constitutive derivative is taken through $\Delta\ubalpha$. Implicit differentiation of \eqref{eq:qres} gives
\begin{equation}
\label{eq:dqimpl}
\frac{\partial q}{\partial q_{\mathrm{trial}}}
=\frac{1}{1+m\beta}
\end{equation}
With $\mathbf{n}=\mathbf{s}^{\mathrm{trial}}/(\sqrt{2}\,q_{\mathrm{trial}})$,
\begin{equation}
\label{eq:dsC}
\frac{\partial\mathbf{s}}{\partial\mathbf{s}^{\mathrm{trial}}}
=\frac{1}{1+\beta}\,\mathbb{I}^{\mathrm{dev}}
+
\left(
\frac{1}{1+m\beta}
-
\frac{1}{1+\beta}
\right)
\mathbf{n}\otimes\mathbf{n}
\end{equation}
The hydrostatic part of $\Sigmaln$ is unaffected by the dashpot. Collecting the volumetric and deviatoric blocks yields
\begin{equation}
\label{eq:Calg}
\begin{aligned}
\frac{\partial\Sigmaln}{\partial(\Delta\ubalpha)}
=\mathbb{C}^{\mathrm{alg}}
&=
\ulambda\,\mathbf{I}\otimes\mathbf{I}
+
\frac{2\umu}{1+\beta}\,\mathbb{I}^{\mathrm{dev}}
+
2\umu
\left(
\frac{1}{1+m\beta}
-
\frac{1}{1+\beta}
\right)
\mathbf{n}\otimes\mathbf{n}\\
&=
\ulambda\,\mathbf{I}\otimes\mathbf{I}
+
\frac{2\umu}{1+\beta}\bigl(\mathbb{I}^{\mathrm{dev}}-\mathbf{n}\otimes\mathbf{n}\bigr)
+
\frac{2\umu}{1+m\beta}\,\mathbf{n}\otimes\mathbf{n}
\end{aligned}
\end{equation}
Hence $\partial\Sigmaln/\partial\C=\mathbb{C}^{\mathrm{alg}}:\partial(\Delta\ubalpha)/\partial\C$. The first term of $\mathbb{C}^{\mathrm{alg}}$ is the elastic volumetric stiffness. The second is the shear stiffness in the plane orthogonal to $\mathbf{n}$, reduced by the dashpot. The third is the consistent modulus along $\mathbf{n}$; it coincides with the plane term for a linear dashpot ($m=1$), in which case the shear modulus is scaled by $1/(1+\beta)=1/(1+\Delta t/t_{\mathrm{M}})$ from \eqref{eq:explicit}. At vanishing trial, $q_{\mathrm{trial}}=0$ and $m>1$, one has $\beta=0$ and $\mathbb{C}^{\mathrm{alg}}=\mathbb{C}^e$. The same formulae hold when $q$ is taken from \eqref{eq:cardan}. Inserting \eqref{eq:Calg} into \eqref{eq:dSdC} gives
\begin{equation}
\label{eq:dSdCfull}
\frac{\partial\mathbf{S}}{\partial\C}
=
\frac{\partial\mathbb{A}^{-1}}{\partial\C}:\Sigmaln
+
\mathbb{A}^{-1}(\C):\mathbb{C}^{\mathrm{alg}}:
\frac{\partial(\Delta\ubalpha)}{\partial\C}
\end{equation}
The spatial algorithmic moduli of the finite-element residual follow by the standard push-forward of $\mathbf{S}$ and of \eqref{eq:dSdCfull}.

\section{Range of validity of the additive logarithmic formulation}
\label{sec:validity}

\begin{figure}[htb!]
\centering
\includegraphics[width=0.52\textwidth]{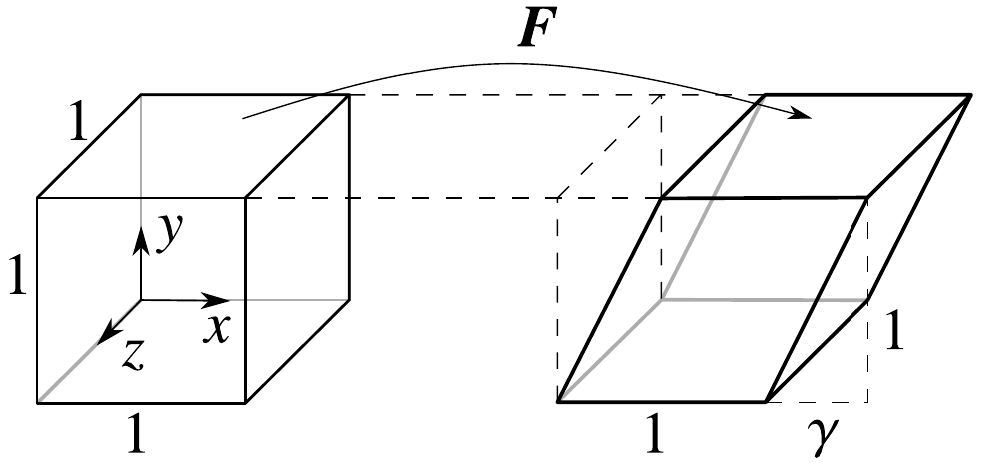}
\caption{Homogeneous simple shear of a unit cube: the deformation gradient \eqref{eq:Fshear} maps the reference square in the $(X_1,X_2)$-plane to a parallelogram of unchanged height, with $\det\F=1$.}
\label{fig:simple-shear}
\end{figure}

Although the formulation is thermodynamically consistent, and although the additive logarithmic structure has been used successfully in finite-strain viscoplasticity --- in particular in the shell models of \citet{sansour1998,sansour2001cmame,sansour2003} and in the anisotropic computational framework of \citet{miehe2002} --- models of this class are known to have limitations. An additive split of a generalised strain of $\C$ ceases to coincide with a multiplicative split of $\F$ on non-coaxial paths \citep{itskov2004}; a Hencky energy that is rank-one convex in $\F$ need not remain so after an additive logarithmic plastic subtraction \citep{neff2016}; and the same class exhibits stress softening, and a possibly localising response, at excessive non-coaxial strain \citep{friedlein2022}. The path on which those limitations are exhibited is homogeneous simple shear at constant $\dot\gamma$. Following \citet{friedlein2022} (see also \citet{thiel2019}), a unit square in the $(X_1,X_2)$-plane is mapped to a parallelogram of unchanged height (Fig.~\ref{fig:simple-shear}),
\begin{equation}
\label{eq:Fshear}
\F
\;=\;
\mathbf{I}+\gamma\,\mathbf{e}_1\otimes\mathbf{e}_2.
\end{equation}

On that path the stretching $\mathbf{d}$ is independent of $\gamma$, while the Cartesian shear component of $\ubalpha=\ln\mathbf{U}$ is not monotonic and $\|\dot{\ubalpha}\|\to 0$ as $\gamma\to\infty$ \citep{gurtin1983,itskov2004}. A power law that takes $\dot{\ubalpha}$ as its rate therefore unloads while the specimen continues to be sheared: that is the mechanism behind the stress drop reported by \citet{itskov2004} and \citet{friedlein2022}.

\begin{figure}[htb!]
\centering
\includegraphics[width=\textwidth]{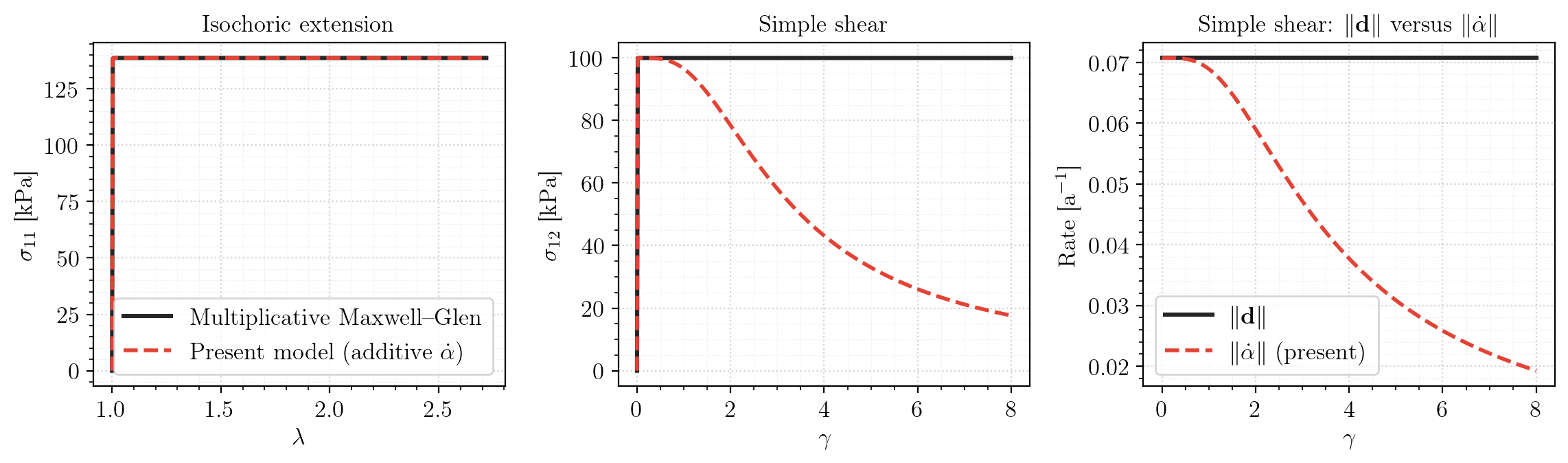}
\caption{Material-point comparison of the present additive logarithmic Maxwell--Glen model with a multiplicative Maxwell--Glen reference, ice moduli, $m=3$. Left: isochoric extension. Centre: homogeneous simple shear \eqref{eq:Fshear} at $\dot\gamma=0.1\,\mathrm{a}^{-1}$. Right: on the shear path, $\|\mathbf{d}\|$ is independent of $\gamma$ while $\|\dot{\ubalpha}\|$ collapses.}
\label{fig:validity-shear}
\end{figure}

The present model splits the logarithmic rate rather than a logarithmic strain. Whether that split inherits the same unloading is therefore checked on \eqref{eq:Fshear}, against a multiplicative Maxwell--Glen model with the same $(A,m,\umu,\kappa)$. Ice moduli are used throughout. Along isochoric coaxial extension, a path on which the two splits coincide, the Cauchy stresses occupy the spatial Glen plateau to machine precision (Fig.~\ref{fig:validity-shear}, left). Along \eqref{eq:Fshear} at $\dot\gamma=0.1\,\mathrm{a}^{-1}$, with $A$ chosen so that spatial Glen holds $\sigma_{12}=100\,\mathrm{kPa}$, the multiplicative reference remains on that plateau, while the present model unloads: the drop is $3\%$ at $\gamma=1$, $22\%$ at $\gamma=2$, and more than $80\%$ at $\gamma=8$ (Fig.~\ref{fig:validity-shear}, centre). The rate seen by the dashpot collapses, although $\|\mathbf{d}\|$ is constant (Fig.~\ref{fig:validity-shear}, right). The rate split therefore belongs to the class diagnosed above.

\begin{figure}[htb!]
\centering
\includegraphics[width=\textwidth]{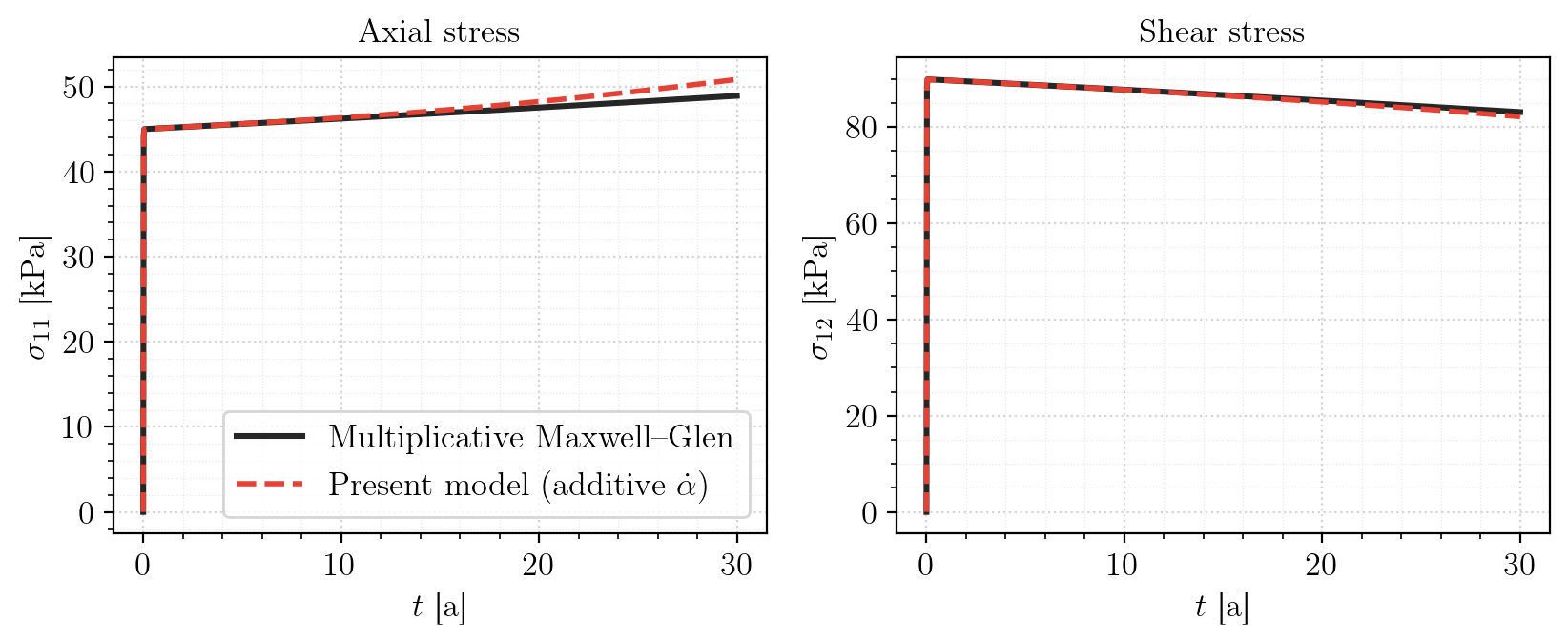}
\caption{Material-point imitation of the shelf trajectory \eqref{eq:Fshelf}, ice moduli, $m=3$. Left: axial stress. Right: shear stress.}
\label{fig:validity-shelf}
\end{figure}

That unloading is decisive only on paths of the type of \eqref{eq:Fshear}. The ice-shelf calculations of this paper are not of that type: the near-front field is dominated by self-weight compression, along-flow extension and cliff flexure, and the principal axes of $\C$ do not spin as in homogeneous simple shear. A material-point imitation of that trajectory is the isochoric planar extension with a modest superimposed shear
\begin{equation}
\label{eq:Fshelf}
\F
\;=\;
\begin{pmatrix}
\lambda & \gamma & 0 \\
0 & 1 & 0 \\
0 & 0 & \lambda^{-1}
\end{pmatrix},
\qquad
\lambda=\mathrm{e}^{\dot\varepsilon\, t},
\qquad
\gamma=\dot\gamma\, t,
\end{equation}
so that $\det\F=1$. With $\dot\varepsilon=5\times 10^{-3}\,\mathrm{a}^{-1}$, $\dot\gamma=2\times 10^{-2}\,\mathrm{a}^{-1}$ and ice moduli, the present Cauchy stress remains within $0.2\%$ of the multiplicative reference after a decade and under $3\%$ after thirty years, at which time $\lambda\simeq 1.16$ and $\gamma=0.6$ (Fig.~\ref{fig:validity-shelf}).

In that envelope the additive logarithmic Maxwell--Glen model remains a tolerable approximation of spatial Glen. The comparison therefore supports the constitutive choice adopted for the ice-shelf calculations of this paper. Ice-stream shear margins, where $\gamma$ becomes large and the kinematics approach \eqref{eq:Fshear}, lie outside it.


\section{Numerical examples}
\label{sec:examples}

A representative boundary-value problem is considered comprising a viscoelastic column under self-weight, compared with the multiplicative exponential update of \citet{schroeder2026}. 
The problem is solved in a finite-element setting by a Newton scheme at the global and Gauss-point levels using linear hexahedral elements.

\begin{figure}[htb!]
\centering
\includegraphics[width=0.3\linewidth]{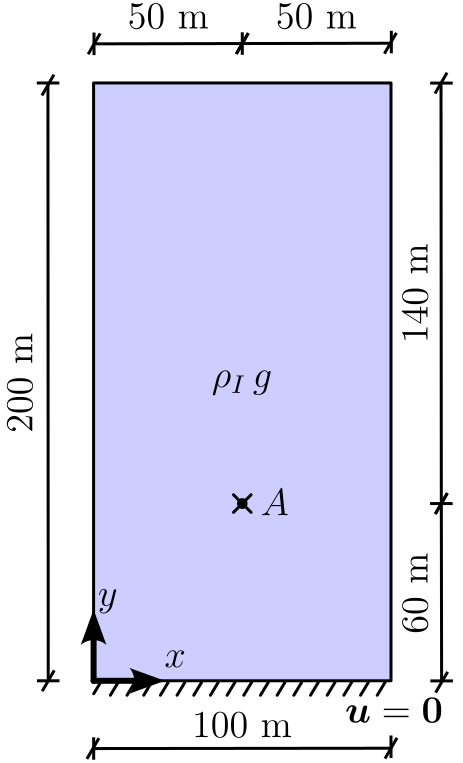}\hfil
\includegraphics[width=0.3\linewidth]{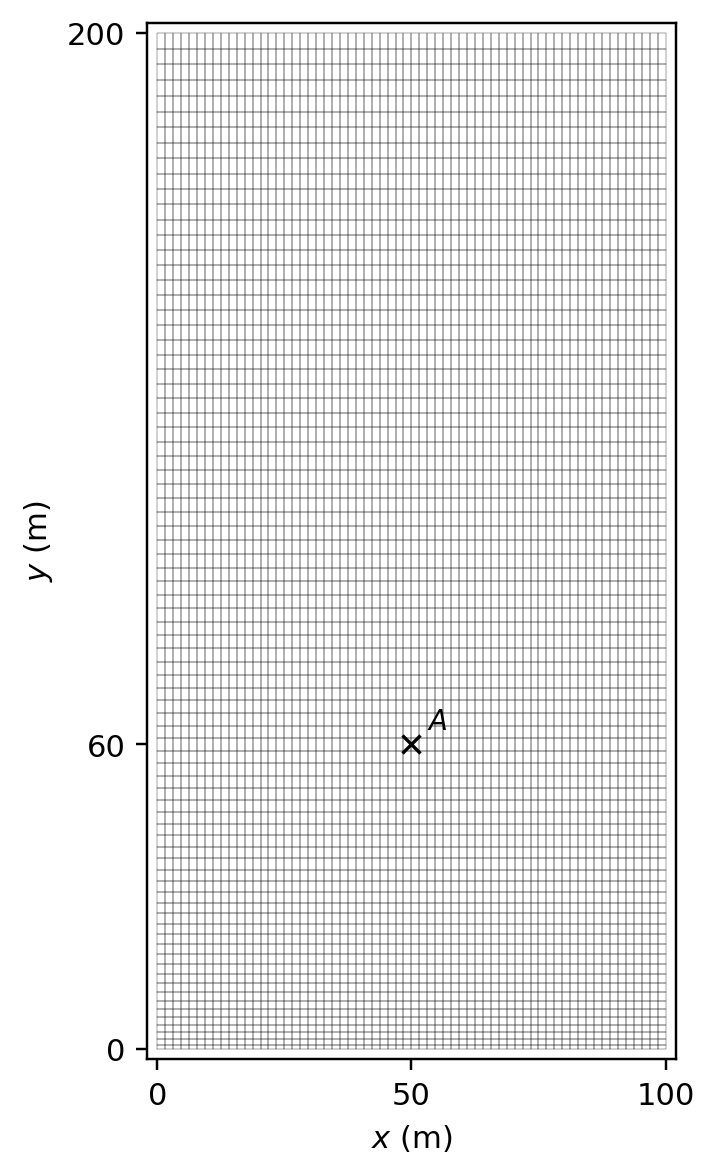}
\caption{Viscoelastic column under self-weight: geometry (left) and mesh (right). The base is fixed; gravity acts downwards. Point $A$ is the control location of \citet{schroeder2026}.}
\label{fig:column}
\end{figure}

At $m=1$, the Glen-type dashpot \eqref{eq:glenlog} reduces to a linear Maxwell element upon the identification $A=1/(2\ueta)$. That reduction is compared with the multiplicative Maxwell of \citet{schroeder2026} on their column (Fig.~\ref{fig:column}).

It consists of a rectangle of width $100\,\mathrm{m}$ and height $200\,\mathrm{m}$, loaded by self-weight, with the material parameters of Table~\ref{tab:column} and $\ueta=10^{14}\,\mathrm{Pa}\,\mathrm{s}$. The base is fixed in both directions; the remaining edges are free of traction. The domain is discretised in plane strain with $64\times 80$ bilinear quadrilaterals, graded towards the base (Fig.~\ref{fig:column}). Point $A$ at $(50,60)\,\mathrm{m}$ is the control location of the reference study. The hold uses $\Delta t=5\,\mathrm{d}$ up to $t=1.5\,\mathrm{years}$.

\begin{table}[width=.9\linewidth,cols=2,pos=htb]
\caption{Material parameters of the column benchmark, after \citet{schroeder2026}.}
\label{tab:column}
\begin{tabular*}{\tblwidth}{@{}LC@{}}
\toprule
Parameter & Value \\
\midrule
Young's modulus $E$ & $9\,\mathrm{GPa}$ \\
Poisson's ratio $\nu$ & $0.325$ \\
Density $\rho$ & $910\,\mathrm{kg}\,\mathrm{m}^{-3}$ \\
Gravity $g$ & $9.81\,\mathrm{m}\,\mathrm{s}^{-2}$ \\
Viscosity $\ueta$ & $10^{14}\,\mathrm{Pa}\,\mathrm{s}$ \\
\bottomrule
\end{tabular*}
\end{table}

\begin{figure}[htb!]
\centering
\includegraphics[width=0.98\linewidth]{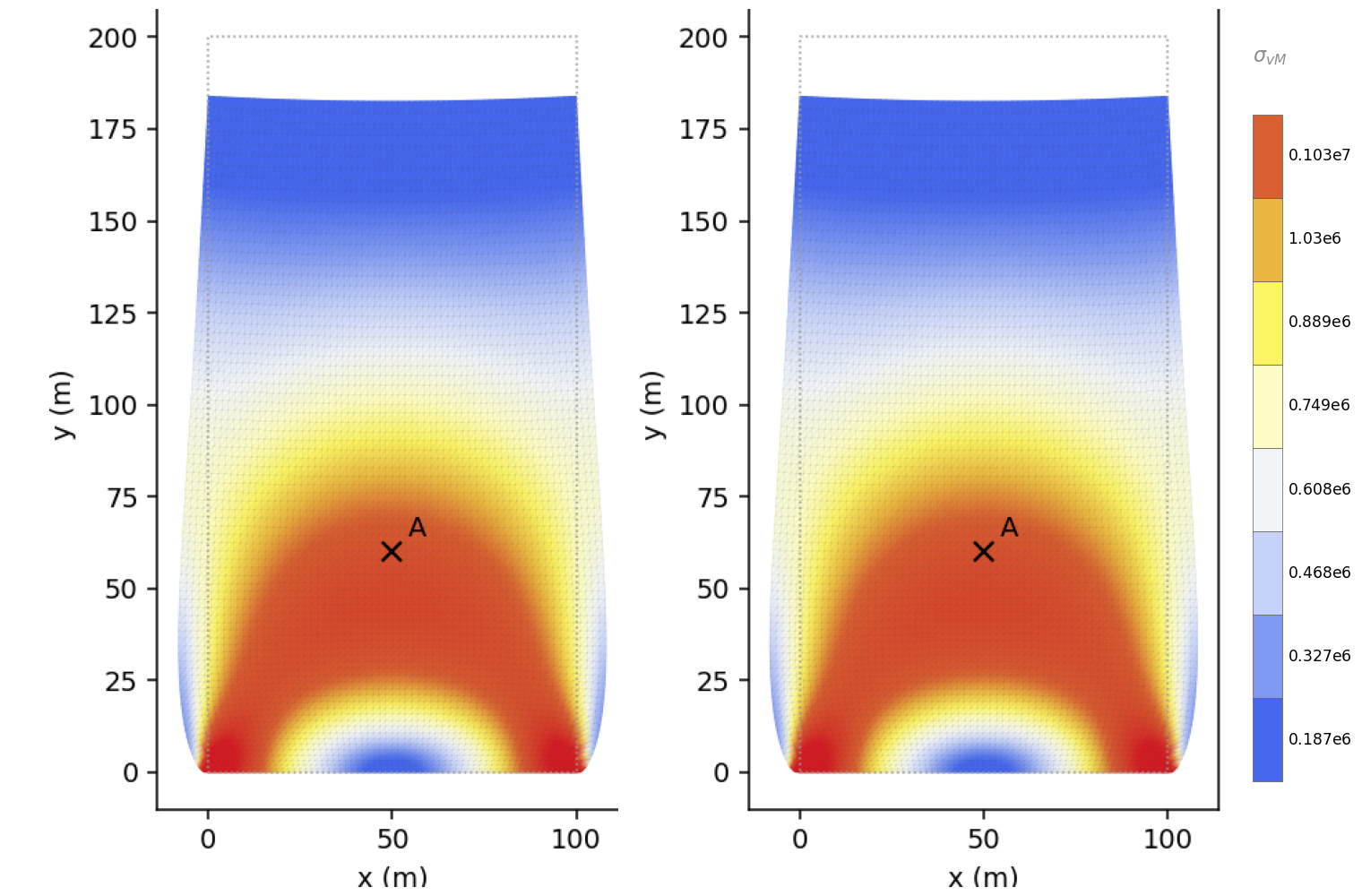}
\caption{Stress distribution $\sigma_{vM}$ in kPa within the column at $t=1.5$ years using the exponential update scheme of \citet{schroeder2026} (left) and the additive logarithmic model (right).}
\label{fig:column-vm}
\end{figure}

Figure~\ref{fig:column-vm} compares the von~Mises stress of the two laws on that mesh, at $t=1.5\,\mathrm{years}$. The top settlement, the lateral bulge and $\sigma_{vM}$ at $A$ agree to the reported digits ($u_y=-17.042$ versus $-17.045\,\mathrm{m}$, $\max|u_x|=8.063$ versus $8.065\,\mathrm{m}$, $\sigma_{vM}(A)=1168.0$ versus $1168.3\,\mathrm{kPa}$; $\lVert\mathbf{u}\rVert_{\infty}$ differs by $3\,\mathrm{cm}$ on an $8\,\mathrm{m}$ bulge): the identification $A=1/(2\ueta)$ reproduces the documented structural creep of \citet{schroeder2026}.

On this path the additive split remains close to the multiplicative Maxwell, consistent with the globally coaxial character of the motion. The lateral bulge takes the eigenframe of $\C$ out of the reference configuration, but $\Sigmaln$ typically rotates with it: the commutator remains small, and the two laws stay close. At $A$ the two $\sigma_{vM}$ fields differ by $0.3\,\mathrm{kPa}$; the element-averaged maxima are $4.37$ and $4.29\,\mathrm{MPa}$.

\begin{figure}[htb!]
\centering
\includegraphics[width=0.88\linewidth]{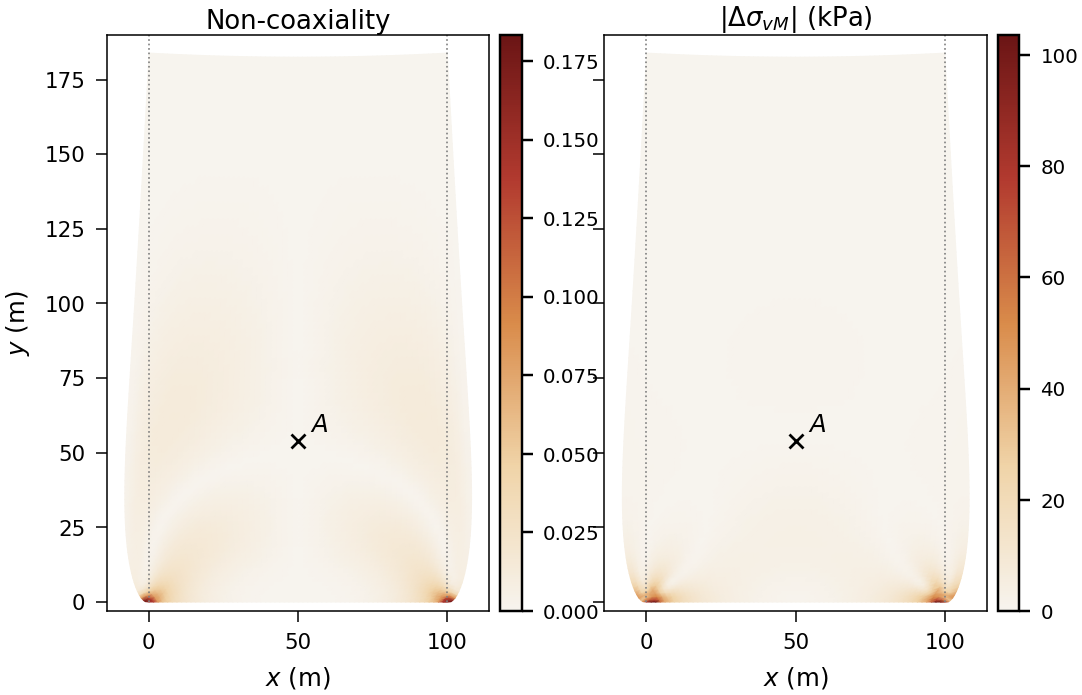}
\caption{Constitutive non-coaxiality $\lVert[\C,\Sigmaln]\rVert/(\lVert\C\rVert\,\lVert\Sigmaln\rVert)$ of the additive model (left) and pointwise difference $\lvert\sigma_{vM}^{\mathrm{add}}-\sigma_{vM}\rvert$ in kPa against \citet{schroeder2026} (right), on the deformed configuration at $t=1.5$ years. The two maps coincide. The residual sits on the kinematic singularity of the fixed corners, already present in the multiplicative description.}
\label{fig:column-coax}
\end{figure}

Figure~\ref{fig:column-coax} shows the constitutive non-coaxiality of the additive model, $\lVert[\C,\Sigmaln]\rVert/(\lVert\C\rVert\,\lVert\Sigmaln\rVert)$, next to the pointwise difference $\lvert\sigma_{vM}^{\mathrm{add}}-\sigma_{vM}\rvert$ against \citet{schroeder2026}. The two maps coincide.

The residual sits on the kinematic singularity of the fixed corners, already present in the multiplicative description, where a Dirichlet condition meets a traction-free edge. There the Gauss maxima are $8.72$ and $8.02\,\mathrm{MPa}$ ($0.70\,\mathrm{MPa}$, about $8\%$). That relative maximum is confined to the strongest geometric singularity. Restricted to the most stressed $1\%$ of the elements ($51$ elements; Table~\ref{tab:column-gap}), the mean and median relative gaps in $\sigma_{vM}$ fall to $1.67\%$ and $1.56\%$ ($34$ and $30\,\mathrm{kPa}$), and those in the equivalent Hencky strain $\varepsilon_{\mathrm{eq}}=\sqrt{2/3}\,\lVert\dev\ubalpha\rVert$ to $0.48\%$ and $0.34\%$. 

\begin{table}[width=.95\linewidth,cols=4,pos=htb]
\caption{Relative discrepancy of the additive model against \citet{schroeder2026} on the most stressed elements (element-averaged fields; mean / median / maximum).}
\label{tab:column-gap}
\begin{tabular*}{\tblwidth}{@{\extracolsep{\fill}}lccc@{}}
\toprule
Zone & $n$ & $\Delta\sigma_{vM}$ & $\Delta\varepsilon_{\mathrm{eq}}$ \\
\midrule
top $0.5\%$ ($1.82$--$4.37\,\mathrm{MPa}$) & $25$ & $2.17$ / $1.98$ / $4.85\%$ & $0.69$ / $0.61$ / $2.22\%$ \\
top $1\%$ & $51$ & $1.67$ / $1.56$ / $4.85\%$ & $0.48$ / $0.34$ / $2.22\%$ \\
\bottomrule
\end{tabular*}
\end{table}

In terms of the diagnostic of Section~\ref{sec:validity}, the lateral bulge corresponds to an effective rotation of order $\max|u_x|/H\simeq 8.06/200\simeq 0.04\,\mathrm{rad}$, roughly an order of magnitude below the shear range ($\gamma\gtrsim 1$) at which the additive split begins to depart measurably from the multiplicative reference (Fig.~\ref{fig:validity-shear}, centre). On this path the two laws are therefore expected to remain close, except at the feet. 

Despite the corner discrepancy, the residual non-coaxiality is weak outside those singularities, and the two descriptions remain close on the stress and strain fields of this example. The comparison verifies the discrete scheme and the $m=1$ reduction of \eqref{eq:glenlog} on the published column of \citet{schroeder2026}.

\section{Conclusions}
\label{sec:conc}
A finite-strain viscoelastic constitutive formulation for glacier ice has been developed in logarithmic strain space. The model retains the familiar structure of a Maxwell element by additively decomposing the material rate of Hencky strain into elastic and viscous contributions, while finite kinematics are retained through the logarithmic strain measure and its work-conjugate stress. The elastic branch is described by isotropic linear elasticity and the viscous branch by a Glen-type power law. The latter reduces to Glen's flow law in the infinitesimal-strain limit and is formulated as a purely deviatoric flow rule, so that viscous incompressibility is satisfied directly. This construction is particularly suited to ice-shelf problems in which elastic strains remain small but finite deformation arises from the accumulation of viscous strain and the progressive evolution of the geometry.

The formulation also leads to a comparatively compact constitutive update. Midpoint evaluation of the finite-strain kinematics combined with an implicit viscous correction reduces the local Maxwell-Glen update to a single scalar nonlinear equation for the equivalent deviatoric stress. For a linear dashpot, the scheme recovers the standard backward-Euler Maxwell update, while for the conventional Glen exponent m=3 the local problem admits a unique solution. The corresponding algorithmic tangent can be derived systematically within the same logarithmic framework, facilitating implementation in an implicit finite-element setting.

An additional advantage of the proposed formulation is that it is entirely rate-based. Both the kinematic decomposition and the Glen-type constitutive relation are expressed in terms of rates, without the need to introduce and evolve a separate viscous deformation gradient. This is particularly attractive for glaciological modelling frameworks, which are themselves typically formulated in terms of velocities and strain rates through Glen's flow law. The logarithmic Maxwell formulation can therefore be incorporated into such rate-based computational settings while retaining finite-strain kinematics and an explicit elastic response. This also provides a convenient basis for extending existing viscous ice-flow models towards viscoelastic descriptions when shorter-timescale loading processes, such as tidal flexure, ocean-wave forcing or rapid changes in calving-front traction, are of interest.

The self-weighted column benchmark confirms the numerical implementation and the reduction to linear Maxwell viscoelasticity. After 1.5 years, the additive logarithmic formulation reproduces the displacements and stress field of the multiplicative reference solution to close agreement: at the control point, the von Mises stresses differ by only 0.3 kPa, while differences over the most highly stressed elements remain small apart from the local singularities at the fixed corners. The comparison therefore demonstrates that, for deformation paths that remain predominantly coaxial, the additive formulation reproduces the structural response of the established multiplicative treatment without requiring a viscous deformation gradient or an intermediate configuration.

The limits of this simplification are nevertheless important. Under homogeneous simple shear, the logarithmic rate decreases as accumulated shear increases, causing the additive model to unload relative to the multiplicative Maxwell-Glen reference. The discrepancy reaches approximately 3\% at $\gamma = 1$, 22\% at $\gamma = 2$, and exceeds 80\% at $\gamma = 8$. By contrast, for a deformation path representative of an ice shelf, combining extension with moderate shear, the difference in Cauchy stress remains below 0.2\% after ten years and below 3\% after thirty years. The proposed formulation should therefore not be regarded as a general replacement for multiplicative finite viscoelasticity, particularly in regions of sustained large shear such as ice-stream shear margins. Its range of applicability is instead the predominantly extensional and flexural deformation characteristic of the ice-shelf and calving-front problems considered here.

\section*{Acknowledgements}

The authors gratefully acknowledge the support of the South African National Antarctic Programme (SANAP). The research work received support from the National Research Foundation (NRF) of South Africa (UID SANAP23042095601). The opinions and conclusions expressed in the research are solely those of the authors and should not be attributed to NRF.

\printcredits

\bibliographystyle{cas-model2-names}
\bibliography{igsrefs}

\end{document}